\documentclass[sigconf]{acmart}
\AtBeginDocument{%
  \providecommand\BibTeX{{%
    \normalfont B\kern-0.5em{\scshape i\kern-0.25em b}\kern-0.8em\TeX}}}

\setcopyright{ccbysa}
\copyrightyear{2023}
\acmYear{2023}
\acmDOI{} 

\acmConference[IMI Workshop'23]{Intelligent Music Interfaces: When Interactive Assistance and Augmentation Meet Musical Instruments}{March 12,
  2023}{Glasgow, UK}
\acmBooktitle{Intelligent Music Interfaces: When Interactive Assistance and Augmentation Meet Musical Instruments, March 12, 2023, Glasgow, UK}

\begin{document}

\title{Can a virtual conductor create its own interpretation of a music orchestra?}

\author{Marc-Philipp Funk}
\email{marc.funk@tum.de}
\affiliation{%
  \institution{Technical University Munich}
  \city{Munich}
  \state{Bavaria}
  \country{Germany}
}

\author{N. Chloe Eghtebas}
\email{eghtebas@in.tum.de}
\affiliation{%
  \institution{Technical University Munich}
  \city{Munich}
  \state{Bavaria}
  \country{Germany}
}


\begin{abstract}
Having a computer do the work for you has become more and more common over time. But in the entertainment area, where a human is a creator, we want to avoid having too much influence on technology. Inspiration on the other hand is still important and as such, we developed a virtual conductor which can generate an emotionally associated interpretation of known music work. This was done by surveying a set number of people to determine, which emotions were associated with a specific interpretation and instruments. As a result of machine learning this conductor was then able to achieve his goal. Unlike earlier studies of virtual conductors, which would replace the role of a human conductor, this new one is supposed to be an assisting tool for conductors. As a result, it will be easier to start on a new interpretation, because it streamlines research time, and provides a technical perspective that can inspire new ideas. By using this technology as a supplement to human creativity, we can create richer, more nuanced interpretations of musical works.
  
\end{abstract}

\ccsdesc[300]{Human-centered computing~HCI theory, concepts and models}

\keywords{machine learning, human computer interaction}

\begin{teaserfigure}
  \includegraphics[width=\textwidth]{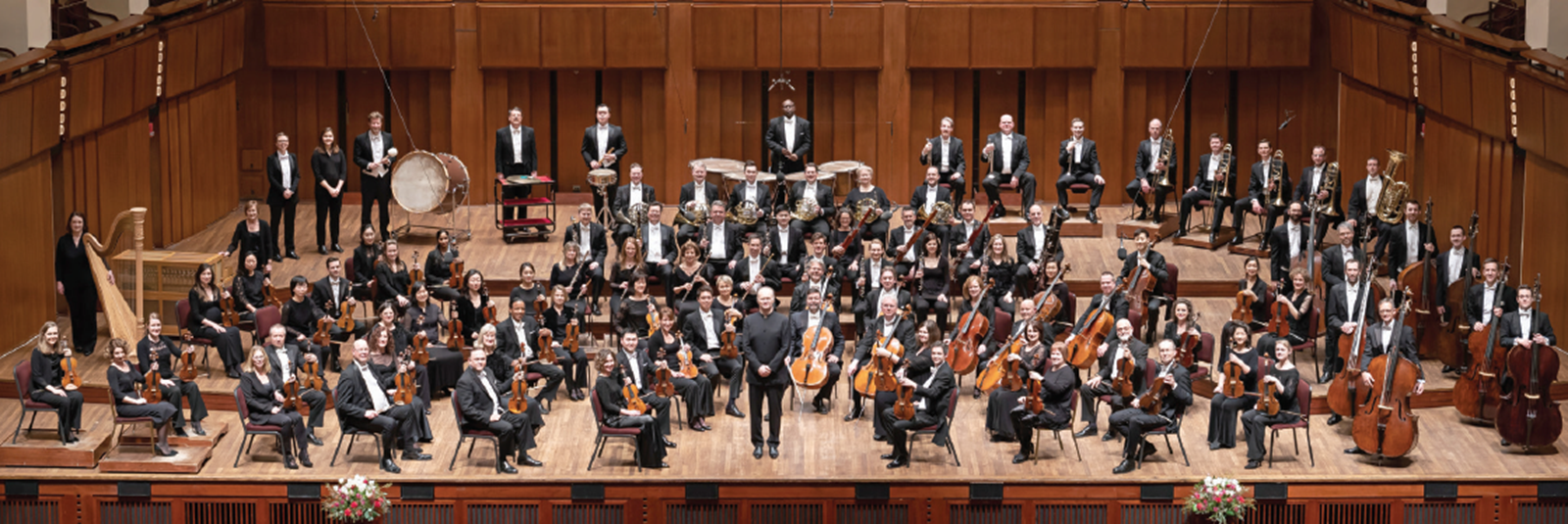}
  \caption{The national symphony orchestra (\url{https://bit.ly/3X8tjl7})}
  \Description{A picture of the national symphony orchestra before beginning their performance.}
  \label{fig:teaser}
\end{teaserfigure}

\maketitle

\section{Introduction}
Today's society enjoys many different forms of entertainment. Two of its most popular kinds are Television and Video Games, which prominently appeared in the 20th century as civilization progressed both technologically and digitally. While the first two rely heavily on computer technology, the second most popular one, music, has a longer history and is less reliant on digitization. One example was classical music, which is often performed by orchestras. Per definition, an orchestra is an ensemble of typically four types of instruments led by a conductor, who leads the instruments thereby conveying his interpretation of a musical work. Past research already introduced human-computer interaction by developing a virtual conductor leading an orchestra. And now this paper investigated if this virtual conductor could generate its interpretation of a past music piece. This was achieved with the help of machine learning, which taught the virtual conductor to differentiate and associate one of the base emotions to a presented music performance. Therefore it was necessary to have a survey group sample example performances with emotions and to let the conductor analyze these findings. As a result, it could generate its interpretation of a known music piece based on a specific emotion and existing work. This was done by making changes to a music piece, which clarified the desired emotion more clearly. As an example, the resulting interpretation expectantly favored one instrument group over another, as they are the main reason for how a performance makes the listener feel. By using this virtual conductor as a supplement to human creativity, conductors can learn how to make changes to a piece to achieve a specific emotional effect. This research opens up new possibilities for the intersection of technology and music, enhancing the experience of classical music for listeners and performers alike.

\section{Related Work}
\subsection{Virtual Conductor}
Past works have devoted themselves to developing a virtual conductor, whose goal was to lead the musicians in place of the human conductor\cite{bos2006interacting}. It was able to conduct different meters and tempos as well as tempo
variations and at the same time was also able to interact with the human musicians by making tempo changes and correcting their mistakes \cite{nijholt2008virtual}. These works played a big part in this paper's motivation. As music is still a medium to bring human imagination to fruition, we didn't want to replace the main part of an orchestra, the conductor and musicians, but we rather focused on supporting the way to the performance. So we concluded to research a system, which generated an interpretation allowing for the collaboration between technology and human creativity.

\subsection{Emotion in music}
Music has the ability to evoke a range of emotions in the listener. But as these emotions are activities of the brain, they are also subjective and vary from person to person. And yet part of it is a consequence of the acoustical differences in music, thereby giving logic to the evocation of emotion while listening \cite{koelsch2018investigating}. Musical parameters such as tempo, harmony, and rhythm also play a role in the perceived emotion of a piece of music. Therefore music emotion recognition systems were designed, which basis and results also helped with generating the emotionally associated work \cite{kim2010music}. To address the subjectivity of emotions, the experiment described in this paper was conducted with a larger sample size and based on the majority decision of participants.

\subsection{Generating music}
Automatic generation of music is still a goal, that is being researched today. Many of the recent AI-based generation services use deep learning models, which have shown promising results. For music past research has derived mathematical models, such as a set of stochastic rules, from a set of musical examples. As such it was able to recognize musical structures and use regularity, inference, and prediction of music to generate its results \cite{dubnov2003using}. The music itself always has a structure \cite{meyer1993sound}, from where sequence likelihoods could be deduced. However, these systems still face challenges in generating high-quality music that is comparable to human compositions. The most important requirement for music generation was hereby a data set and training methods \cite{sturm2019machine}, which was also the basis for this research.

\section{Research Questions/Hypotheses}
Based on accomplishments from previous works, we now set out to find an answer to the following question: Was it possible and viable to let a virtual conductor machine learn how to newly generate an emotional music sheet interpretation for a specific existing music work? In this context emotional meant, that this new score was supposed to have changes from the original music based on the selected emotion the resulting interpretation should convey. Through machine learning with the data of the experiment, we wanted to teach the virtual conductor to know how to associate emotion by analyzing a score. The main focus lay on the possibility to let it create an interpretation after it knew what emotions change in a music piece. As such, we investigated, if the generation tool was a viable way for learning conductors to get inspired by how to convey a specific emotion with a specific work. 

\section{Prototype/Implementation}
The implementation of the generating system relied on the achievements provided by the past works mentioned above. But we took these accomplishments and results, adapted and expanded on them, all to put them to use in a new context and environment to bring forth this new research. Therefore all this was accomplished by using already existing hardware. That was most important for the machine learning aspect, which learned and collected data about patterns in given music scores\cite{7993626}. The used model for this was the supervised machine learning model "Linear Regression" \cite{montgomery2021introduction}. For that we made use of a music score analysis system to read out all the necessary information and digitize the data \cite{576283}. 
\subsection{Learning the logic behind emotion}
Through the experiment specified below we acquired some data from past orchestra performances, where the conductor made adjustments to the work, which resulted in his interpretation. As the experiment data itself was not nearly enough, other publicly published performances, as well as their reviews were taken into consideration. Special weight was given to the intention of the responsible conductor if it was known. The required parameters needed for teaching the system were the sheet of the interpretation, the original score, and the survey results of the associated emotion of parts of as well as the whole performance. Based on all this, it was possible to implement the machine learning algorithm. Firstly the differing sheets were compared by the system with a comparison algorithm. After analyzing them, these scores existed as data, where the necessary information could be obtained digitally in. Therefore it was possible to also compare the data of two music pieces. Every important information about the performance was in the score, such as when an instrument ensemble started, their volume, tempo, etc. As a result, even more information was won than just by listening to a performance, which could be difficult to identify even for experienced musicians. These differences as well as the most prevalent emotion for the specific differences between the score provided the virtual conductor with the necessary information to connect the score with emotion. After that, it saved these results. Linear Regression Analysis was able to identify the relationship between the variable of interest, the emotions, and the input parameters. With that, the system learns to predict emotions based on the given data, which is an important requirement for the second part. Lastly connecting these most common similarities to the emotional intention taught the virtual conductor about the patterns and logic of how to achieve the sensation in the listener.

\begin{figure*}
  \includegraphics[width=\textwidth]{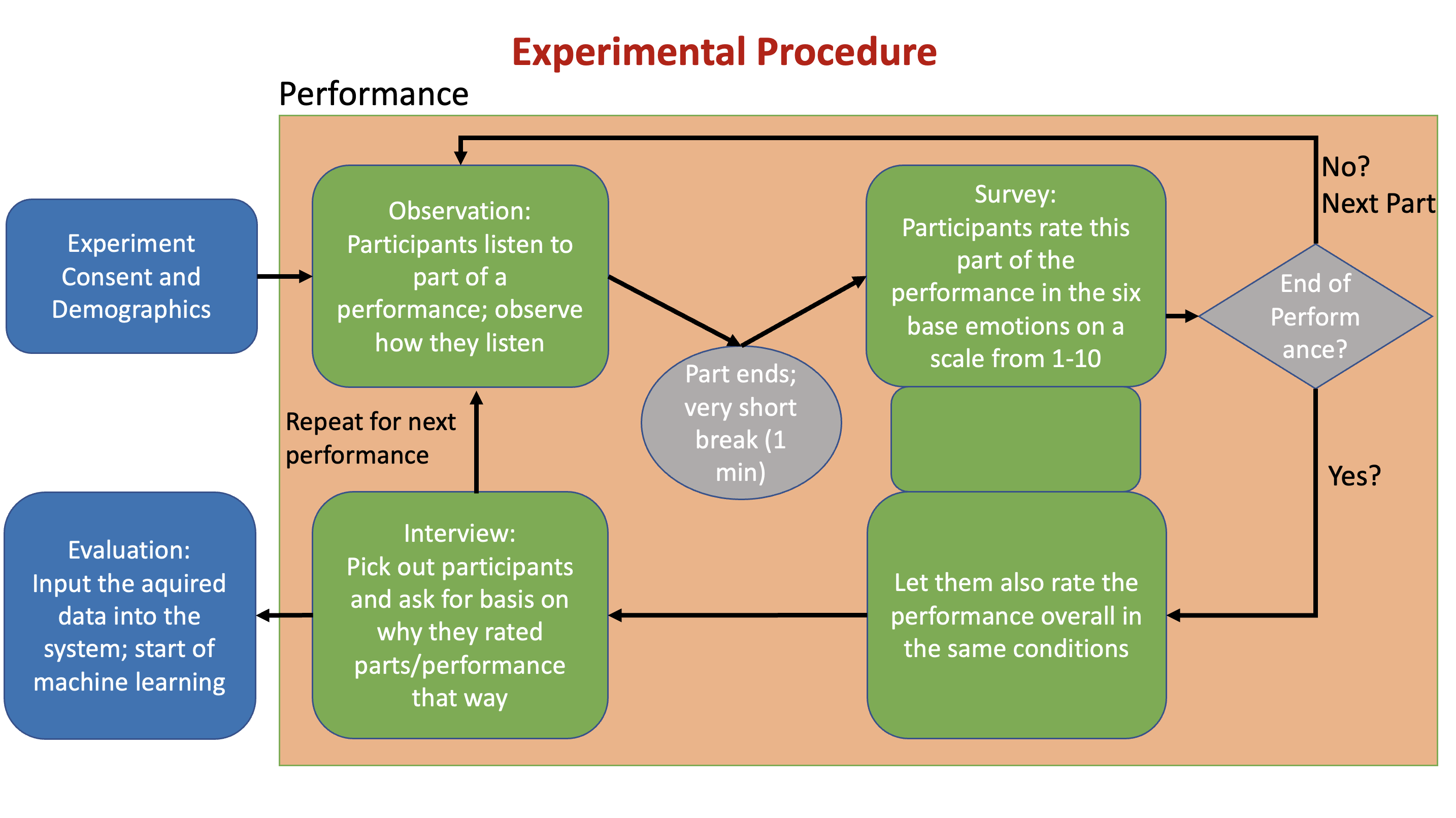}
  \caption{Experimental Procedure}
  \Description{The graphical representation of the research experiment}
\end{figure*}

\subsection{Interpretation Generation}
After gathering sufficient data through machine learning, the next step of the research process followed. The goal of the system was to make changes to a music sheet as such it counted as an interpretation. The input parameters for the generation were the original music sheet and the desired emotion for the interpretation out of the base emotions, which were both requested by a user in an interaction with the system. In some cases, it was also important to process older interpretations of the same work, if that was not already done in the first step. The reason for that is, that the generation algorithm relied on analyzing these past performances. Similar to before a pattern search was performed to find the specific parts of the performance, where changes by the conductor were made to convey his intentions. With this acquired information the virtual conductor was now able to apply its changes near and at parts with often made adaptations in other music works, if these existed. If was that not the case, the system would change up parts, where related instruments played. But as it mattered which sensation the resulting piece of music wanted to convey, the logic from Linear Regression Analysis in the first part was now used. With its predictions, it knew, which changes could be done to achieve the sensation. As a result, the associated instruments were given a more prevalent part overall, whereas instrument groups on the other side of the emotional spectrum were now more in the background. Something like that was achieved by adjusting their volumes and note length as an example. This as well was done by making changes to the data of the file with the help of music composition systems \cite{miranda2001composing}. Re-contextualizing this information on other music works produced a unique result, which already counted as an interpretation. As a means to make these generated pieces more valid, this algorithm also had to be trained. Therefore it was a good opportunity to use the virtual conductor or a virtual orchestra itself to try to conduct and play the new interpretation as a way to identify faults, such as parts that were impossible to play. To evaluate its success the same experimental procedure, which is clarified below, had to be repeated. Over time it grew into a valid tool by being able to bring inspiration and ideas to conductors.

\section{Research Method}
To bring a system, that was supposed to help people, to fruition also required the support of people. On top of that, this research dealt with the complexity of human emotions, which could not be easily transferred into such a program, that didn't know emotions, to begin with. Emotions are known for being subjective, which is why one person's opinion on the conductor's intention of an interpretation was biased and therefore flawed data to teach the system. As a result, more than one opinion was necessary to refine the needed information. The research method made use of that requirement. 
\subsection{The basic emotions}
To narrow down a conductor's intention on an emotional spectrum, we had to set the bounds and restrictions of that range. As such, we utilized the ''Circle of the Circumplex'' model \cite{10.3389/fpsyg.2019.00781}, where every emotion could be located on an axis based on the parameters. These were joy and sadness, as well as anticipation and surprise, both being opposites of one another. But to get further information from the experiment, we also included anger and calmness in addition to fear and confidence. These were the possible sensation, which an interpretation was rated for.
\subsection{Experimental Procedure}
A set number of around 30 people was invited to take part in this experiment, after getting their approval. The experiment they participated in took one day overall. First, they were all shown past publicly available orchestra performances of a conductor's interpretation. These were split into different parts with very short breaks in between. The listeners' task was to rate these parts on a scale of 1 to 10 for every base emotion. They were also encouraged to note down the reasons for their rating. The parts were based on the four movements of a symphony \cite{stein1962structure}. If there were major instrument changes in these movements, they were also broken down into more sections. While the participants listened, the experiment management also observed them for any notable listening behavior, which could be affiliated with emotion. It's important to mention that the music piece was played without any additional information, such as the name of the piece or the emotion it was supposed to convey, to prevent any biases. After a whole piece, the overall performance was also graded in the same way. Lastly in an interview afterward, some of the participants were asked about the performance they were shown before, and they explicated their specific reasons for their opinion more accurately, with their notes as a basis. The people for the interview were picked based on if there was an earlier mentioned behavior during the performance and based on a good mix of the participant categories.\\
The results confirmed that certain instrument groups could be connected to certain emotions. As such string instruments brought forth more sad, percussion instruments happy, woodwind instruments neutral, and brass instruments fearful emotions. As was already mentioned a very similarly structured experiment had to be carried out for the synthesized performances of the generated interpretations. As this was a more continuous process to find faults, fewer participants were necessary.

\subsection{Participants}
This group consisted of people with a non-musician background and some with a musical background with at least 5 years of experience in music theory or playing a related instrument. Regular orchestra visitors, who could be one or the other, were included as well. A mix of these categories of partakers was essential, as each of them had different viewpoints to consider. A non-musician assessed mostly based on the feelings he had while listening, without taking technical aspects of the music into account. On the other hand, a musician was able to undermine why a specific part felt this way, because of volume, instruments, or harmony. The regular orchestra visitor has experienced many different interpretations of works. That is the reason, why they were more likely to grasp a conductor's intention with specific parts of a music piece. It is important to consider, that while the musician usually performs better in recognizing musical structures, non-musicians did almost just as well in discriminating between instrument timbres, which is the different sound of an instrument. Additionally, musicians and non-musicians would perform equally poorly in recognizing identical chords played on different instruments \cite{beal1985skill}. But as the focus lay more on the emotional tone color, that aspect was not needed for the procedure. Ultimately, gathering input from a diverse group of participants helped to ensure that the system accurately reflected the range of emotions conveyed by a conductor's interpretation.

\section{Conclusion and future work}
In this research paper, we proposed a music interpretation generator, which made changes to an existing music piece for an orchestra to convey a specific emotion. This was mainly accomplished with the help of machine learning. Unlike other HCI systems in the music area, this one was meant to be a supporting tool, bringing new ideas and inspiration to conductors. A significant consequence is that the orchestra field might open up to a broader audience, which could result in more upcoming conductors. For future work, this research can be expanded upon by letting the resulting interpretation get conducted by a virtual conductor as well, where new feedback from the system to the system can be acquired, making it ideal for machine learning. As such the virtual conductor comes close to a human conductor in terms of its tasks, although it's important to mention again, that it is not supposed to replace the latter. Another possibility is to let it also get played by a real orchestra, where it can be played for the first time by real musicians, whose feedback is invaluable as well. Both these ways would also further improve this research, as through continuous feedback the generation only grows more reliable. This system has the potential to expand into other musical genres and could also be utilized for music therapy, with a significant impact that warrants further investigation. Overall, the proposed system has the potential to open up new avenues for creativity and innovation in the music industry, and it will be interesting to see how it will develop in the future.

\bibliographystyle{ACM-Reference-Format}
\bibliography{sample-base}

\appendix

\end{document}